\documentclass{elsart}

\usepackage{graphicx}

\usepackage{amssymb}

\begin{document}

\begin{frontmatter}



\title{Detectability of the Supernova Relic Neutrinos
	and Neutrino Oscillation}


\author[label1]{S. Ando},
\ead{ando@utap.phys.s.u-tokyo.ac.jp}
\author[label1,label2]{K. Sato}, and
\author[label3,label4]{T. Totani}

\address[label1]{Department of Physics, University of Tokyo,\\
7-3-1 Hongo, Bunkyo, Tokyo 113-0033, Japan}
\address[label2]{Research Center for the Early Universe, 
University of Tokyo,\\
7-3-1 Hongo, Bunkyo, Tokyo 113-0033, Japan}
\address[label3]{Princeton University Observatory, Peyton Hall, Princeton,
NJ 08544, USA}
\address[label4]{Theory Division, National Astronomical Observatory,\\
Mitaka, Tokyo, 181-8588, Japan}

\begin{abstract}
We investigate the flux and the event rate 
of the supernova relic neutrino background (SRN)
at the SuperKamiokande detector 
for various neutrino oscillation models with parameters inferred
from recent experimental results.
A realistic model of neutrino emission from supernova explosions 
and several models of the cosmic star formation history are
adopted in the calculation.
The number flux over entire energy range is found to be
$11-15 ~\mathrm{cm^{-2}s^{-1}}$.
We discuss the detection possibility of SRN at SuperKamiokande, 
comparing this SRN flux
with other background neutrinos in more detail than previous studies.
Even though there is no energy window in which SRN is dominant, 
we might detect it as the distortion of the other background event.
We found in the energy range $17-25~\mathrm{MeV}$ the expected event rate 
at SuperKamiokande $0.4-0.8 ~\mathrm{yr^{-1}}$.
In this range, ten-year observation might enable us to detect
SRN signal (at one sigma level) in the case of LMA solar 
neutrino solution.
We also investigate event rate at SNO and KamLAND.
Although we can find energy window, the expected event rate is rather
small (0.03 yr$^{-1}$ for SNO, 0.1 yr$^{-1}$ for KamLAND).

\end{abstract}

\end{frontmatter}

\newpage

\section{Introduction}
\label{sec:Introduction}

A core-collapse supernova explosion produces a number of neutrinos and
99\% of the gravitational energy is transformed to neutrinos.  It is
generally believed that the core-collapse supernova explosions have
traced the star formation history in the universe and have emitted a
great number of neutrinos, which should make a diffuse background.
This supernova relic neutrino (SRN) background is one of the targets
of the currently working large neutrino detectors, SuperKamiokande
(SK) and Sudbury Neutrino Observatory (SNO).  Comparing the predicted
SRN spectrum with the observations by these detectors provides us
potentially valuable information on the nature of neutrinos as well as
the star formation history in the universe.  This SRN background has
been discussed in a number of previous papers
\cite{rf:Bisnovatyi-Kogan}-\cite{rf:Steigman}.  The work after Totani et
al. \cite{rf:Totani_1} takes into account the realistic star formation
history inferred from various observations and theoretical modeling of
galaxy formation, to calculate the SRN flux.  Totani et
al.\cite{rf:Totani_1} calculated the energy flux of SRN at SK detector
and compared it with neutrinos emitted by other sources (solar,
atmospheric, and reactor neutrinos and so on).  Then they concluded
that the visible event rate of SRN at SK is 1.2 yr$^{-1}$ in the
energy range from 15 to 40 MeV.  On the other hand, Kaplinghat et
al. \cite{rf:Steigman} calculated the upper limit of SRN and discussed
the possibility of the detection at SK detector comparing to the other
backgrounds.  Their result is that at the energy range from 15 to 40
MeV, where no significant background had been considered to exist in
previous studies, there is a huge background of the invisible muon
decay and the detection of SRN is difficult.  They also discussed the
effects of neutrino oscillation and briefly mentioned that {\it if}
the SRN flux is in the vicinity of their upper bound and all three
flavors are maximally mixed, it may be detectable as a distortion of
the expected muon background.

In this paper we calculate the SRN flux and the event rate at SK, and discuss
about the detectability of SRN, with the following new aspects compared with
previous studies: 1) realistic neutrino oscillation parameters are
incorporated based on the recent solar and atmospheric neutrino experiments, 2)
a realistic neutrino spectrum from one supernova explosion is used, which is
obtained from a numerical simulation by the Lawrence Livermore group, and 3)
we have examined other contaminating background events against the
detection of SRN, in more detail than previous studies.

Recent experiments of SK and SNO support neutrino oscillation by
solar \cite{rf:Fukuda_1}-\cite{rf:SNO} and atmospheric neutrino
\cite{rf:Fukuda_2} data.  Although some previous studies ({\it e.g.}
\cite{rf:Steigman}) mentioned the possibility of neutrino
oscillation, no quantitative calculation of SRN has been made
incorporating the realistic oscillation parameters.
Here we consider
four neutrino oscillation models, which satisfy the solar and
atmospheric neutrino data, and investigate the dependence of the
positron spectra at SK on the oscillation models. 
If neutrino oscillation occurs,
$\bar{\nu}_{\mu,\tau}$'s are converted into $\bar{\nu}_e$'s which are
mainly detected at SK detector.  Because $\bar{\nu}_{\mu,\tau}$'s
interact with matter only through the neutral-current reactions in
supernovae, they are weakly coupled with matter compared to
$\bar{\nu}_e$.  Thus the neutrino sphere of $\bar{\nu}_{\mu,\tau}$'s
is deeper in the core than $\bar{\nu}_e$'s and their temperatures are
higher than $\bar{\nu}_e$'s.  Therefore neutrino oscillation enhances
the mean $\bar{\nu}_e$'s energy and enhances event rate at SK
detector.  

In all of the past studies, neutrinos emitted from a supernova are assumed to
obey the Fermi-Dirac distribution with zero chemical potential. However, since
neutrinos are not in the thermal equilibrium states in supernovae, the real
spectrum should be different from the pure Fermi-Dirac distribution.  Thus we
use in this paper a realistic supernova model established by the Lawrence
Livermore group \cite{rf:Wilson} which in fact shows clear difference from the
Fermi-Dirac distribution \cite{rf:Totani_3}.
 
There are several background events which hinder the 
detection of SRN.  These includes
atmospheric and solar neutrinos, anti-neutrinos from
nuclear reactors, and decay electrons from invisible muons.  We should
find the energy region which is not contaminated by these background
events and then calculate the detectable event rate of SRN.  
By careful examination of these events, we found that there is a narrow
energy window at which the detection of the SRN event might be
possible.

In addition we use three models of the cosmic star formation history,
i.e., the evolution of star formation rate (SFR) density of high redshift
galaxies (e.g., \cite{rf:Madau_1}), which is inferred from
several star formation indicators such as rest-frame UV luminosity,
H$\alpha$ lines, or dust emission in far-infrared or submillimeter
wavebands. It should be noted that
SFR models contain various uncertainties, especially at high
redshift regions. ({\it e.g.}, the faint ends of luminosity functions
have not been well established at high redshifts and the uncertainties
in dust extinction are large at all redshifts \cite{rf:Steidel}.) This 
is the reason we applied several SFR evolution models described below.

Throughout this paper, we consider only electron antineutrinos
($\bar{\nu}_e$'s) from collapse-driven supernovae, because
$\bar{\nu}_e$'s are most easily detected in a water \v{C}herenkov
detector like SK.  

This paper is organized as follows: In Section \ref{sec:Models}, we
illustrate the neutrino oscillation models, the supernova model, and
the supernova rate models considered in this paper and discuss about
these models.  Formulations for calculation of flux and event rate at
SK are given in Section \ref{sec:Formulation}.  In Section
\ref{sec:Results}, we show the calculated flux and event rate.
Detailed discussions especially on other
background events are presented in Section \ref{sec:Discussion}.

\section{Models of Neutrinos and Supernova Rate}
\label{sec:Models}

\subsection{Supernova Model}

\begin{figure}[htbp]
\begin{center}
\includegraphics[width=15cm]{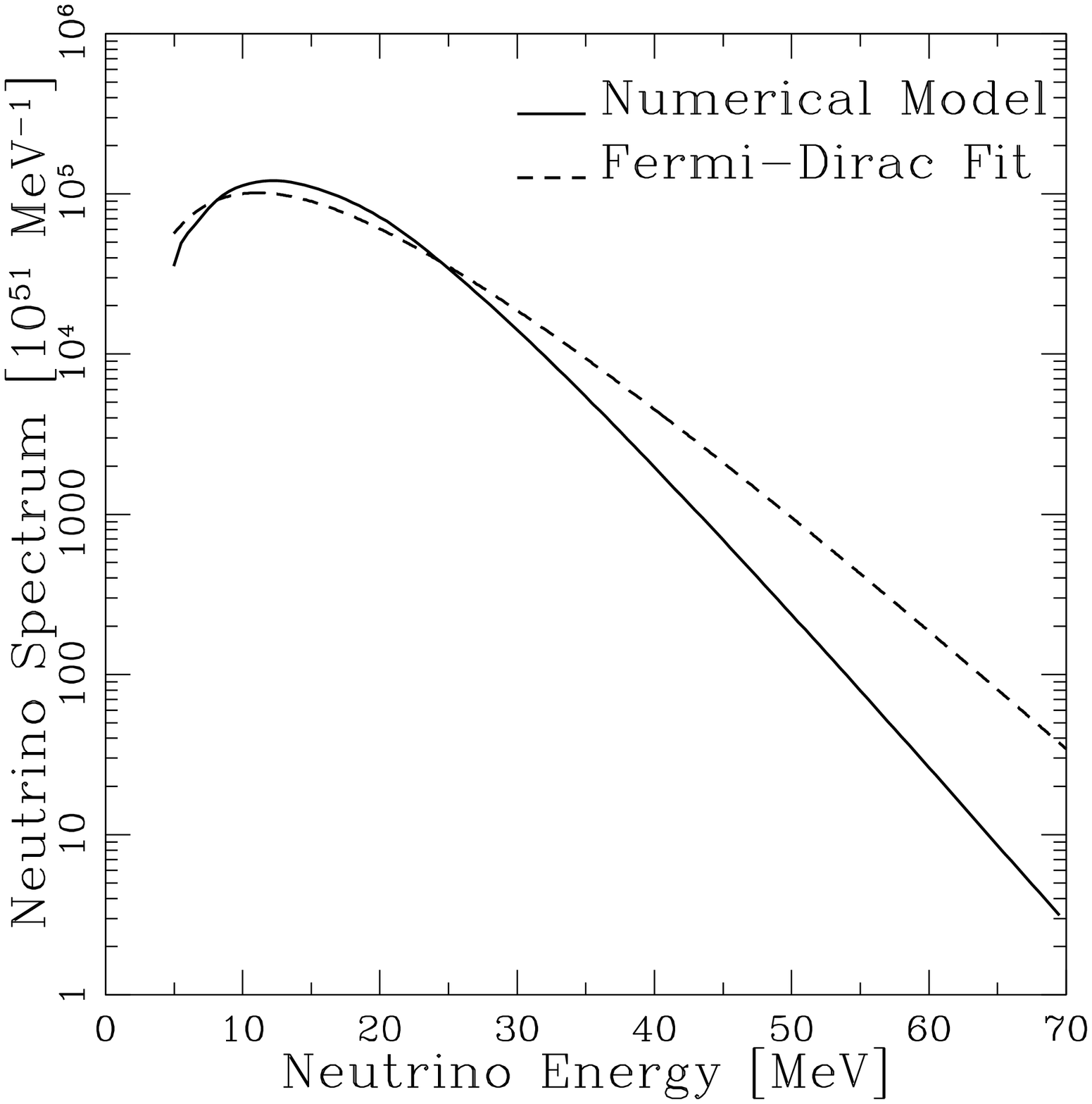}
 \caption{Energy spectrum of $\bar{\nu}_{e}$ of the 
          numerical supernova model used in this paper. The dashed line is the
	  Fermi-Dirac fits which have the same luminosity with the numerical model. The chemical potential is set to zero for the FD distribution.}
 \label{fig:spectrum}
\end{center}
\end{figure}

We use a realistic model of a collapse-driven supernova calculated by
the Lawrence Livermore group \cite{rf:Wilson}.  We show in
Fig. \ref{fig:spectrum} the time-integrated spectrum for $\bar{\nu}_e$'s (e.g.,
see Ref. \cite{rf:Totani_3} for detail).  We also show the Fermi-Dirac
(FD) distribution with zero chemical potential for comparison.
Comparing with the FD model used in the other studies, the deficit of
both low- and high-energy neutrinos can be seen.


In this paper, we assume that this numerical model represents all of
the past supernovae, although in this model the progenitor of the
supernova has been assumed to have the mass of $\sim 20M_{\odot}$.  It
is clearly an oversimplification, but we note that the mean mass
of progenitor stars of type II supernovae (above $8 M_{\odot}$)
weighted by number is
about $\sim 15 M_{\odot}$ when a typical initial mass function is
applied \cite{rf:Woosley}.

Another assumption made in the simulation is the isotropic radiation
of neutrinos.  All supernova progenitors are rotating, and it may have
significant effect on the degree of isotropy of neutrino emission
\cite{rf:Shimizu}, although the effect is very difficult to estimate
quantitatively.  However, since our interest is in the SRN background,
which is the sum of all past supernova neutrinos, the rotation effect
is expected to be small.

\subsection{Neutrino Oscillation Models}
\label{subsec:osci}

We adopt the four parameter sets for the neutrino mixing (see Table
\ref{table:parameter}) and the normal mass hierarchy.  These
parameter sets are introduced to explain the observations of the solar
and the atmospheric neutrinos \cite{rf:Fukuda_1}-\cite{rf:Fukuda_2}.
The neutrino spectra emitted by each supernova are calculated
numerically by Takahashi et al. \cite{rf:Takahashi} for the same
parameter sets assuming the normal mass hierarchy.  (Supernova model
they have used is the same one discussed in the previous subsection.)
We use their calculated $\bar{\nu}_e$ spectrum as that from any 
supernova.

In Table \ref{table:parameter}, ``LMA'' and ``SMA'' indicate MSW
solution of the solar neutrino problem (see Ref. \cite{rf:Kuo}, for
the review of MSW effect).  Recent SK and SNO observations 
\cite{rf:Fukuda_1,rf:SNO} show that LMA solution are more favorable but SMA
solution cannot be rejected at more than  $3\sigma$ level (see
Ref. \cite{rf:Krastev} and references therein).  Therefore we try
the SMA solution as well.  The suffixes ``-L'' and ``-S'' indicate whether
$\theta_{13}$ is large or small.  A large (small) $\theta_{13}$ means
``higher resonance'' is adiabatic (nonadiabatic).  
(There are two resonance points where neutrino mass eigenstates may flip
in the supernova matter, and we call them lower and higher resonance points.
The lower resonance point is at lower density, and the higher resonance point
at higher density.) The adiabatic
higher resonance enhances the energy of the electron neutrinos and
enhances the event rate of $\nu_e$ scattering, but does not affect the
electron anti-neutrinos \cite{rf:Dighe}.  In this paper, since we deal
with only the reaction of $\bar{\nu}_e$'s, this ``-L'' and ``-S''
hardly influence the result.


Now, we semi-qualitatively present a simple illustration of the expected spectral shape of $\bar{\nu}_e$'s for LMA and SMA models.
We can naively deal with anti-neutrino oscillation effect as vacuum oscillation, since $\bar{\nu}_e$'s are not affected by the resonance.
Further we take two generation formalism, for simplicity (this approximation is justified when $\theta_{13}$ is sufficiently small).
The conversion probability from $\bar{\nu}_{\mu}$ to $\bar{\nu}_e$ (and its inverse) averaged over distance is 
\begin{equation}
P(\bar{\nu}_{\mu} \rightarrow \bar{\nu}_e)=P(\bar{\nu}_e \rightarrow \bar{\nu}_{\mu})=\frac{1}{2}\sin^2 2\theta_{12}.
\label{eq:conv_prob}
\end{equation}
This probability depends only on $\theta_{12}$.
Then, $P(\bar{\nu}_{\mu} \rightarrow \bar{\nu}_e)=P(\bar{\nu}_e \rightarrow \bar{\nu}_{\mu})=0.44 ({\rm LMA}),2.5\times10^{-3} ({\rm SMA})$.
Using these conversion probabilities we can present $\bar{\nu}_e$ flux as
\begin{eqnarray}
F_{\bar{\nu}_e}&=&[1-P(\bar{\nu}_e\rightarrow \bar{\nu}_\mu)]F_{\bar{\nu}_e}^0
	+P(\bar{\nu}_\mu\rightarrow \bar{\nu}_e)F_{\bar{\nu}_\mu}^0 \\
	&\simeq&\left\{
	\begin{array}{@{\,}ll}
		0.57F_{\bar{\nu}_e}^0+0.44F_{\bar{\nu}_\mu}^0
			 & \mathrm{(LMA)}\\
		F_{\bar{\nu}_e}^0 & \mathrm{(SMA)}
	\end{array}
	\right. ,
\end{eqnarray}
where $F_{\bar{\nu}}^0$ means neutrino flux for no oscillation model.
This equation shows that in LMA models the $\bar{\nu}_e$ flux at
higher energy region is enhanced, on the other hand that at lower energy
region is suppressed, because the original mean $\bar{\nu}_\mu$'s energy 
is higher than that of $\bar{\nu}_e$'s as mentioned in Section
\ref{sec:Introduction}.

However, in the case of the inverted hierarchy,
this situation changes dramatically although we do not give
calculation for this case.  In this
case, the higher resonance point is in the anti-neutrino sector
\cite{rf:Dighe}, and hence resonant conversion of $\bar{\nu}_e$'s and
$\bar{\nu}_{\mu,\tau}$'s is possible. The conversion probability would be
dependent on the neutrino energy. Another possibility of 
such resonant conversion effect is the spin-flip oscillation of
$\bar{\nu}_e$'s by a flavor-changing neutrino magnetic moment
\cite{rf:totani-sato}.

\subsection{Supernova Rate}

\begin{figure}[htbp]
\begin{center}
\includegraphics[width=15cm]{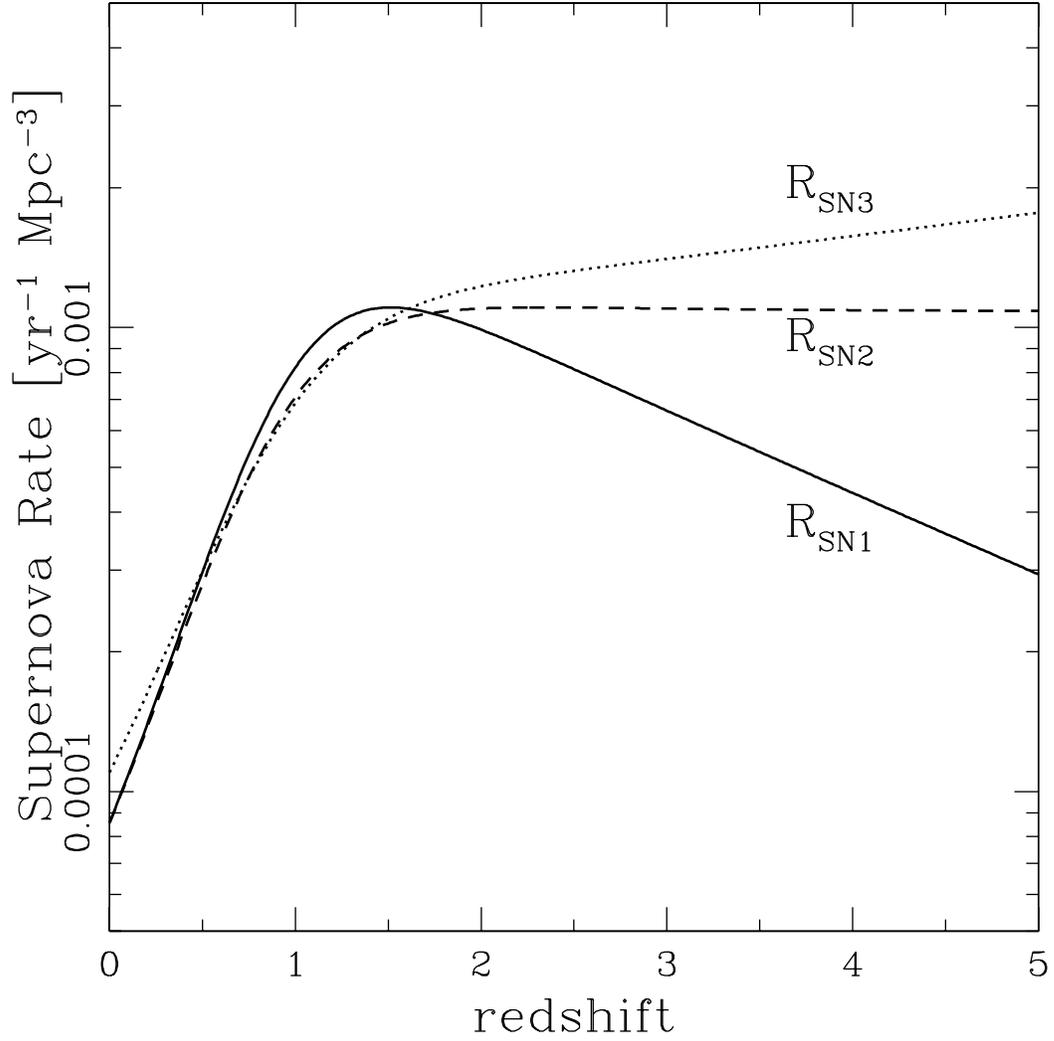}
 \caption{Supernova rate evolution on the cosmological time scale. These lines
 are for a $\Lambda$-dominated cosmology ($\Omega_m=0.3,
 \Omega_{\lambda}=0.7$). The Hubble constant is taken to be 70 km s$^{-1}$
 Mpc$^{-1}$.}  \label{fig:SNrate}
\end{center}
\end{figure}

A number of studies have modeled the expected evolution of the cosmic
SFR with redshift.  The cosmic SFR can now be traced to $z \simeq 4$
observationally, although some details remain controversial.  We use
three SFRs per unit comoving volume which are also used in
Ref. \cite{rf:Madau_2}, where all of them are derived following the
same method employed by Madau et al. \cite{rf:Madau_1} first.  They
use the intergalactic absorption to identify high redshift galaxies in
broadband multicolor survey, and convert the UV luminosity to SFR and the
metal ejection rate assuming a typical
initial mass function (IMF) of stars. The supernova rate is expected
to be proportional to SFR since the lifetime of progenitors of 
core-collapse supernovae is much shorter than the cosmological time scale.
We show in Fig. \ref{fig:SNrate} the
supernova rates corresponding to these three SFR models that are
explained below.


In the Einstein-de Sitter (EdS) universe ($\Omega_m=1.0,
\Omega_{\lambda}=0.0$), the first (hereafter SF1) is taken from
Ref. \cite{rf:Madau_3}:
\begin{equation}
R_{SF1}(z) = 0.3 h_{65}\frac{\exp(3.4z)}{\exp(3.8z)+45} M_{\odot}
~\mathrm{yr^{-1}Mpc^{-3}},
\end{equation}
where $h_{65}=H_0/65~$km s$^{-1}$ Mpc$^{-1}$.  This SFR increases
rapidly between $z=0$ and $z=1$, peaks between $z=1$ and $z=2$, and
gently declines at higher redshifts.  This includes an upward
correction for dust reddening of $A_{1500}=1.2$ mag.  
The original SFR (before correction) should be interpreted as lower
limits to the real values, since a significant fraction of UV light
from young stars could be absorbed and re-emitted in far-infrared
band. In the second model, the SFR remains instead roughly constant
at $z \gtrsim 2$ (SF2) \cite{rf:Steidel}, as:
\begin{equation}
R_{SF2}(z) = 0.15 h_{65}\frac{\exp(3.4z)}{\exp(3.4z)+22} M_{\odot}
~\mathrm{yr^{-1}Mpc^{-3}}.
\end{equation}
We should consider SF2 in addition to SF1, because of the uncertainties associated with the incompleteness of the data sets and the amount of dust extinction at early epochs. 
The third SFR (SF3) \cite{rf:Madau_2},
\begin{equation}
R_{SF3}(z) = 0.2 h_{65}\frac{\exp(3.05z-0.4)}{\exp(2.93z)+15} M_{\odot}  ~\mathrm{yr^{-1}Mpc^{-3}}
\end{equation}
represents even more star formation at early epochs.  This SFR is
based on the studies which suggest that the evolution of the SFR up to
$z \approx 1$ may have been overestimated \cite{rf:Cowie}, while the
rates at high-$z$ may have been severely underestimated due to large
amounts of dust extinction \cite{rf:Blain}.

These SFR evolutions should change when a different cosmological
model is assumed, since they are based on the observed data of 
high-$z$ galaxies. The correction to other cosmological models can be
written as (e.g., \cite{rf:totani-yoshii-sato}):
\begin{eqnarray}
R_{SF}(z;\Omega_m,\Omega_{\lambda},h_{65})
	&=&h_{65}\frac{\sqrt{(1+\Omega_mz)(1+z)^2-\Omega_{\lambda}(2z+z^2)}}
		{(1+z)^{3/2}} \nonumber \\ 
	& &{} \times R_{SF}(z;1,0,1).
\label{eq:SF}
\end{eqnarray}

To obtain the supernova rate ($R_{SN}$), we multiply the SFRs by the
coefficient
\begin{equation} 
\frac{\int _8 ^{125} dm \phi (m)}{\int _{0.01} ^{125} dm m \phi (m)}=0.0122M_{\odot}^{-1},
\end{equation}
where $\phi (m)$ is the Salpeter IMF ($\phi(m)\propto m^{-2.35}$) and $m$
is the stellar mass in
solar units \cite{rf:Madau_2}. (Here, we assume that all stars whose
mass is greater than $8M_{\odot}$ explode as core-collapse supernovae.) This
resulting rates agree with the local observed value.  We label these
models ``SN1'', ``SN2'', and ``SN3'', respectively.

\section{Formulation of Flux Calculations}
\label{sec:Formulation}

In section 2, we calculated the number of emitted $\bar{\nu}_e$'s from
a supernova per unit energy $q$, $dN_{\nu}/{dq}$, and supernova rate,
$R_{SN}$.  We remark that $R_{SN}(z)$ is supernova rate per comoving
volume, and hence we should multiply the factor $(1+z)^3$ to obtain
the rate per physical volume at that time.  The present number density
of $\bar{\nu}_e$'s, whose energy is in the interval of the $q \sim
q+dq$, emitted in the interval of the redshift $z \sim z+dz$ is given
by
\begin{eqnarray}
dn_{\nu}(q)&=&R_{SN}(z)(1+z)^3 \frac{dt}{dz} dz \frac{dN_{\nu}((1+z)q)}{dq} (1+z)dq (1+z)^{-3}\\
&=&R_{SN}(z) \frac{dt}{dz} dz \frac{dN_{\nu}((1+z)q)}{dq} (1+z)dq,
\end{eqnarray}
where the factor $(1+z)^{-3}$ comes from the expansion of the universe.
The Friedmann equation gives the relation between $t$ and $z$ as follows:
\begin{equation}
\frac{dz}{dt}=-H_0(1+z)\sqrt{(1+\Omega_mz)(1+z)^2-\Omega_{\lambda}(2z+z^2)}.
\end{equation}

We now obtain the differential number flux of SRN,
$dF_{\nu}(q)/{dq}$, using the relation
$dF_{\nu}(q)/{dq}=c[dn_{\nu}(q)/{dq}]$:
\begin{eqnarray}
\frac{dF_{\nu}}{dq}&=&\frac{c}{H_0}\int _0 ^{z_{\rm max}} 
			R_{SN}(z) \frac{dN_{\nu}((1+z)q)}{dq} \nonumber \\
	& & {} \times \frac{dz}{\sqrt{(1+\Omega_mz)(1+z)^2-
		\Omega_{\lambda}(2z+z^2)}}, 
\label{eq:flux_cal}
\end{eqnarray}
where we assume that gravitational collapses began at the redshift
parameter $z_{\rm max}=5$.

It can be seen that, from equations (\ref{eq:SF}) and
(\ref{eq:flux_cal}), the flux does not depend on the cosmological
parameters such as $\Omega_m, \Omega_{\lambda}, $ and $H_0$, while in
Ref. \cite{rf:Totani_1}, the flux depend on these parameters.  The
illustration of this difference is as follows: Totani et
al. used supernova rate evolution derived from their theoretical
model of galaxy evolution which reproduces various properties of
present-day galaxies. On the
other hand, our supernova rate is based on the observational estimate
of luminosity densities of high-$z$ galaxies, and the dependence
of the cosmological volume element on the parameters such as
$\Omega_m$ and $\Omega_\lambda$ is cancelled out.
Therefore, our supernova rate depends on the
cosmological parameters while the neutrino flux does not depend on them.

\section{Results}
\label{sec:Results}

\subsection{Calculation of SRN Flux}

\begin{figure}[htbp]
\begin{center}
\includegraphics[width=15cm]{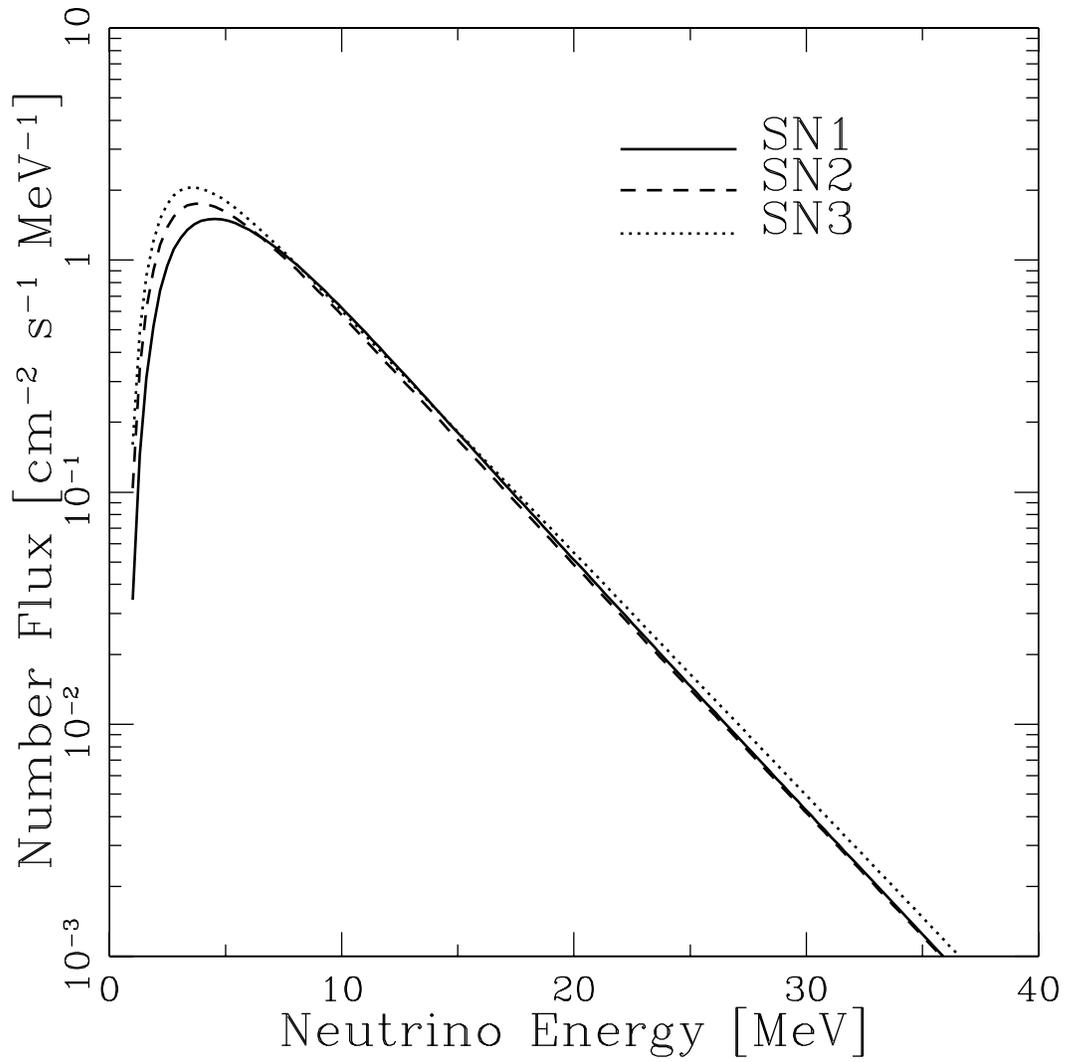}
 \caption{Number flux of $\bar{\nu}_e$'s for the three supernova rate models, assuming ``no oscillation'' case.} 
 \label{fig:flux_SN}
\end{center}
\end{figure}

We calculate the flux of the SRN using the formula
eq. (\ref{eq:flux_cal}) for various models.  In
Fig. \ref{fig:flux_SN}, the flux for the three supernova rate models
are shown assuming no oscillation model.  The fluxes of these
models are almost the same above $\sim 8$ MeV and the models with more
SFR at early epochs have a higher peak at lower neutrino energy.
These properties come from the following effects: The energy of
neutrinos which were emitted at redshift $z$ is reduced by a factor
$(1+z)^{-1}$ when we observe.  Then at observation, the high energy
tail ($\gtrsim 8 ~\mathrm{MeV}$) is mainly contributed by low redshift
supernovae.  Since at low redshift three supernova rate models are
almost the same, the fluxes at $E_{\bar{\nu}_e}\gtrsim8 ~\mathrm{MeV}$
are not much different.  Similarly, because high redshift
supernova neutrinos contribute more at low energy region when we
observe, we can see model dependence at low energy clearly.

\begin{figure}[htbp]
\begin{center}
\includegraphics[width=15cm]{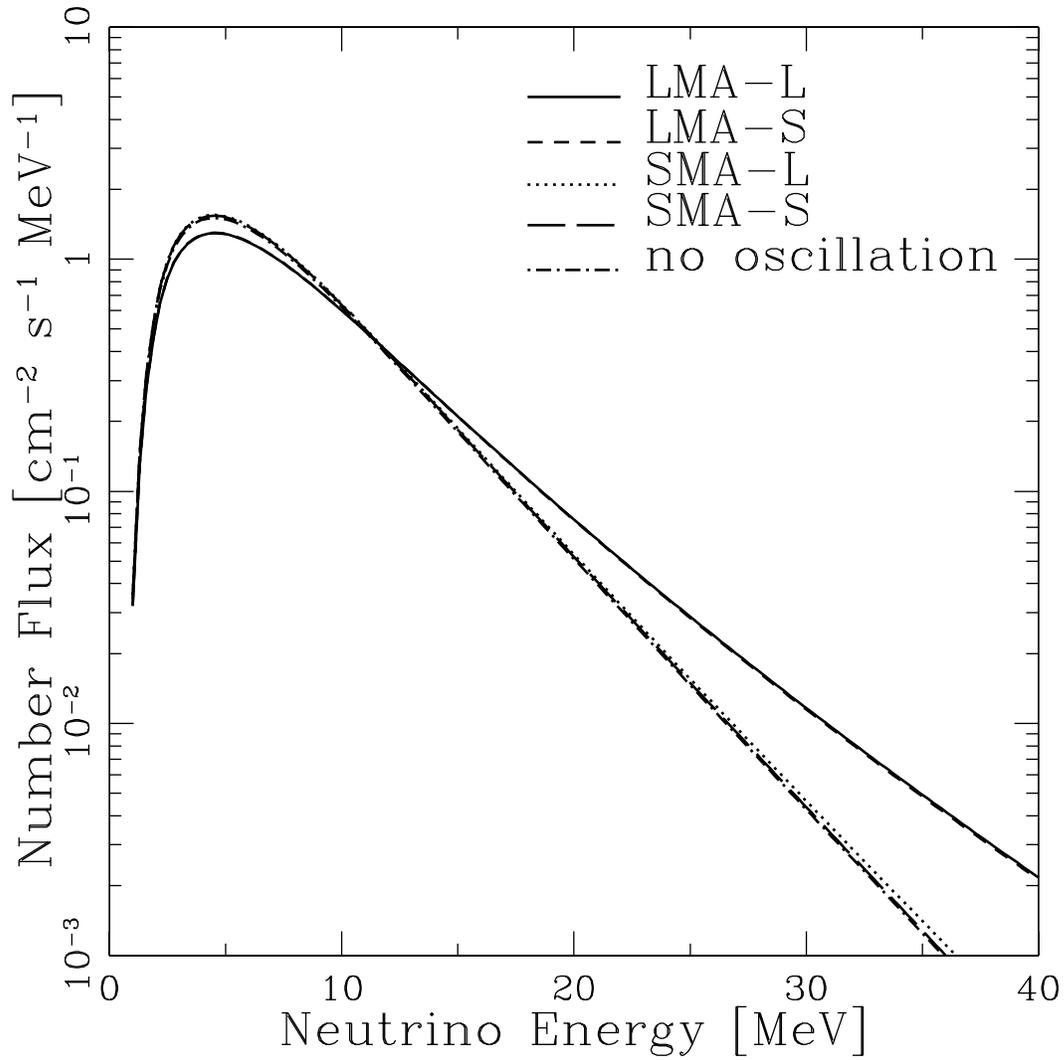}
 \caption{Number flux of $\bar{\nu}_e$'s for the neutrino oscillation
 models. In this figure ``SN1'' model is used for supernova rate
evolution.}  
\label{fig:flux_osci}
\end{center}
\end{figure}

In Fig. \ref{fig:flux_osci}, we show the flux of various neutrino
oscillation models assuming SN1 model.  In SMA and no
oscillation models we can see higher peak around $\sim 5$ MeV than
in LMA models, because in LMA models the low energy
$\bar{\nu}_e$'s are deficit due to the conversion into
$\bar{\nu}_{\mu,\tau}$'s.  But above $\sim 10$ MeV, in LMA models
we can see more flux, because $\bar{\nu}_{\mu,\tau}$'s which have
higher mean energy at production have more changed into
$\bar{\nu}_e$'s than in the case of SMA or no
oscillation. (See also qualitative explanation in Section
\ref{subsec:osci}.)


Integrated flux over the entire neutrino energy range is shown in
Table \ref{table:flux}.


\subsection{Event Rate at SuperKamiokande Detector}

The SK detector is a water \v{C}erenkov detector whose fiducial mass
is 22,500 ton.  The detector efficiency is 100$\%$ for electrons
(positrons) whose energy is above 5 MeV, and 50$\%$ at 4.2 MeV.  We
only consider the reaction $\bar{\nu}_e+p\rightarrow e^++n$, because
the cross section [$9.52\times
10^{-44} (E_e / \mathrm{MeV}) (p_e / \mathrm{MeV}) \mathrm{cm}^2$
\cite{rf:Vogel}] is much larger than the other reactions.

\begin{figure}[htbp]
\begin{center}
\includegraphics[width=15cm]{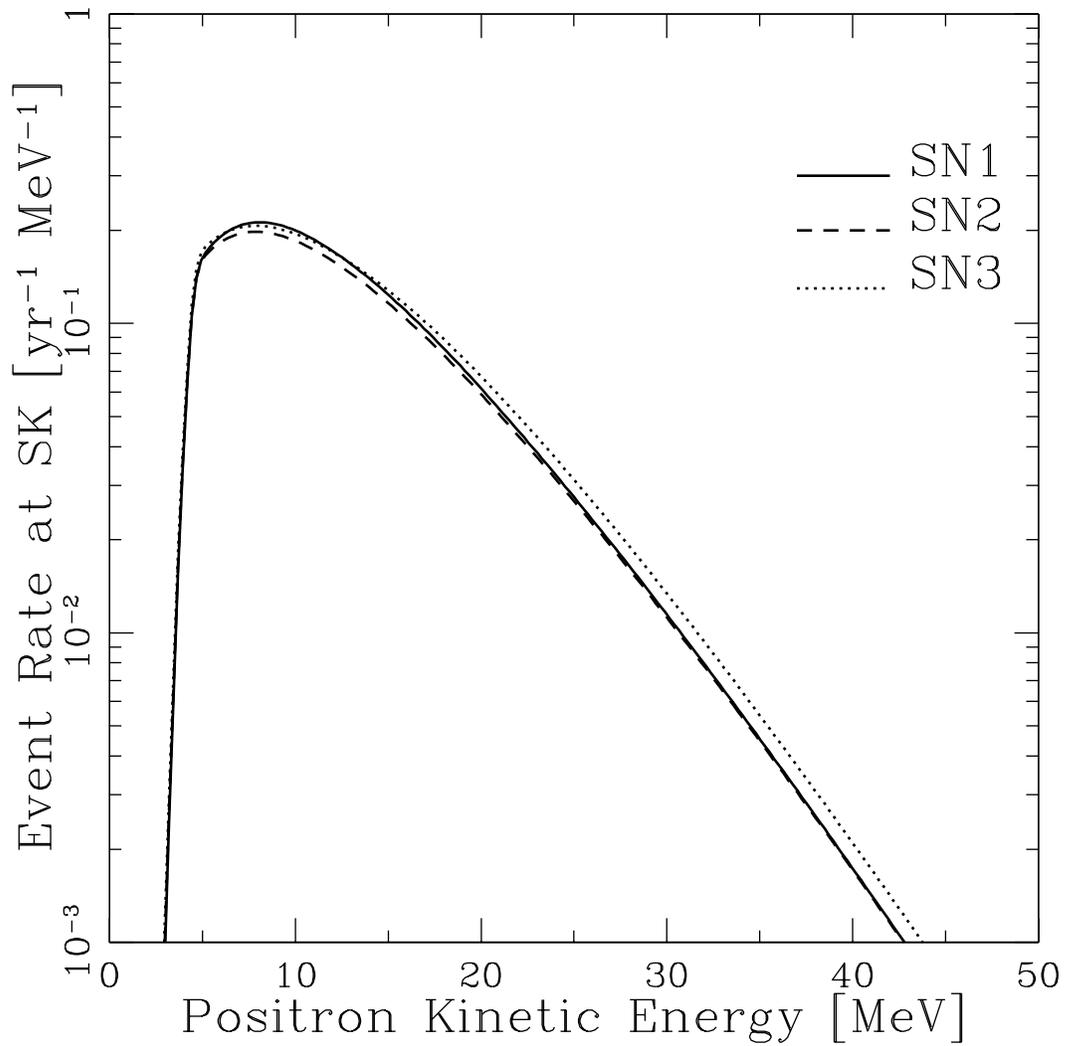}
 \caption{Event rate of $\bar{\nu}_e$'s at SK for the three supernova rate models, assuming ``no oscillation'' case.} 
 \label{fig:event_SN}
\end{center}
\end{figure}

In Fig. \ref{fig:event_SN}, the event rate at SK of the three
supernova rate models are shown assuming no oscillation model.
Because of the detector threshold, we see only the positrons whose
energies are above $\sim 5$ MeV, so that the differences between
models are small as shown also in Fig. \ref{fig:flux_SN} where the
model dependence is also small for neutrinos above $\sim 7$ MeV.
Then, although our three supernova rate models have very different
properties at high redshift regions, our results are not influenced by
the behavior of SFRs at those high $z$ regions.  (These
results are hardly changed when we assume the other LMA or SMA
models, instead of the no oscillation model.)
 
\begin{figure}[htbp]
\begin{center}
\includegraphics[width=15cm]{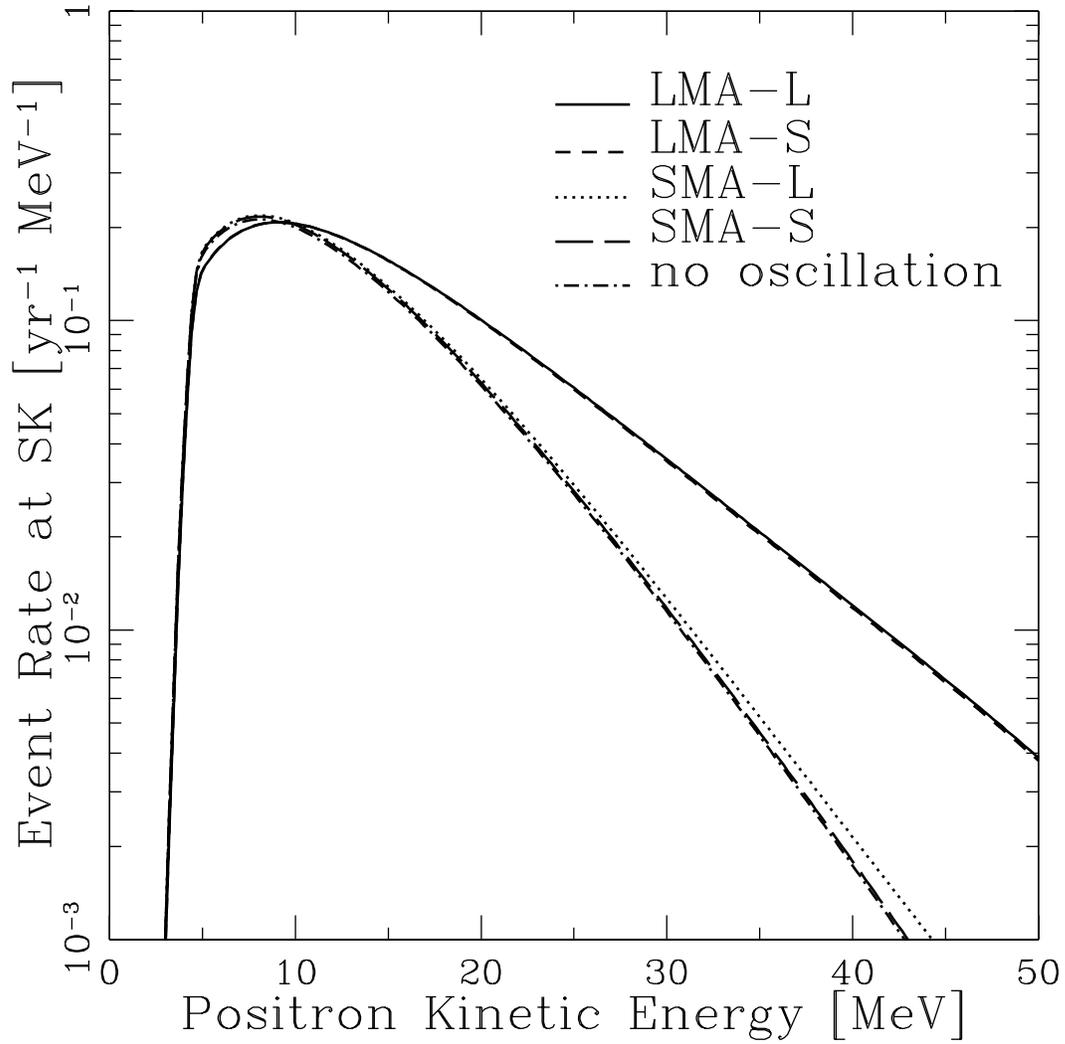}
 \caption{Event rate of $\bar{\nu}_e$'s at SK for the neutrino oscillation
 models. In this figure ``SN1'' model is used for supernova rate evolution.}
\label{fig:event_osci}
\end{center}
\end{figure}

In Fig. \ref{fig:event_osci}, we show the event rate of various
neutrino oscillation models assuming SN1 model.  We can see 
clear difference of the LMA model from SMA or no
oscillation models, especially at the high energy tail.  This
property results from the flux dependence on oscillation models (see
also Fig. \ref{fig:flux_osci}).


As a result, if we can detect SRN events above $\sim10$ MeV, we can
discriminate the LMA model from the SMA or no oscillation models
in any supernova rate models and any sets of cosmological parameters.

\section{Discussion}
\label{sec:Discussion}

\subsection{Background Events against the Detection}

\begin{figure}[htbp]
\begin{center}
\includegraphics[width=15cm]{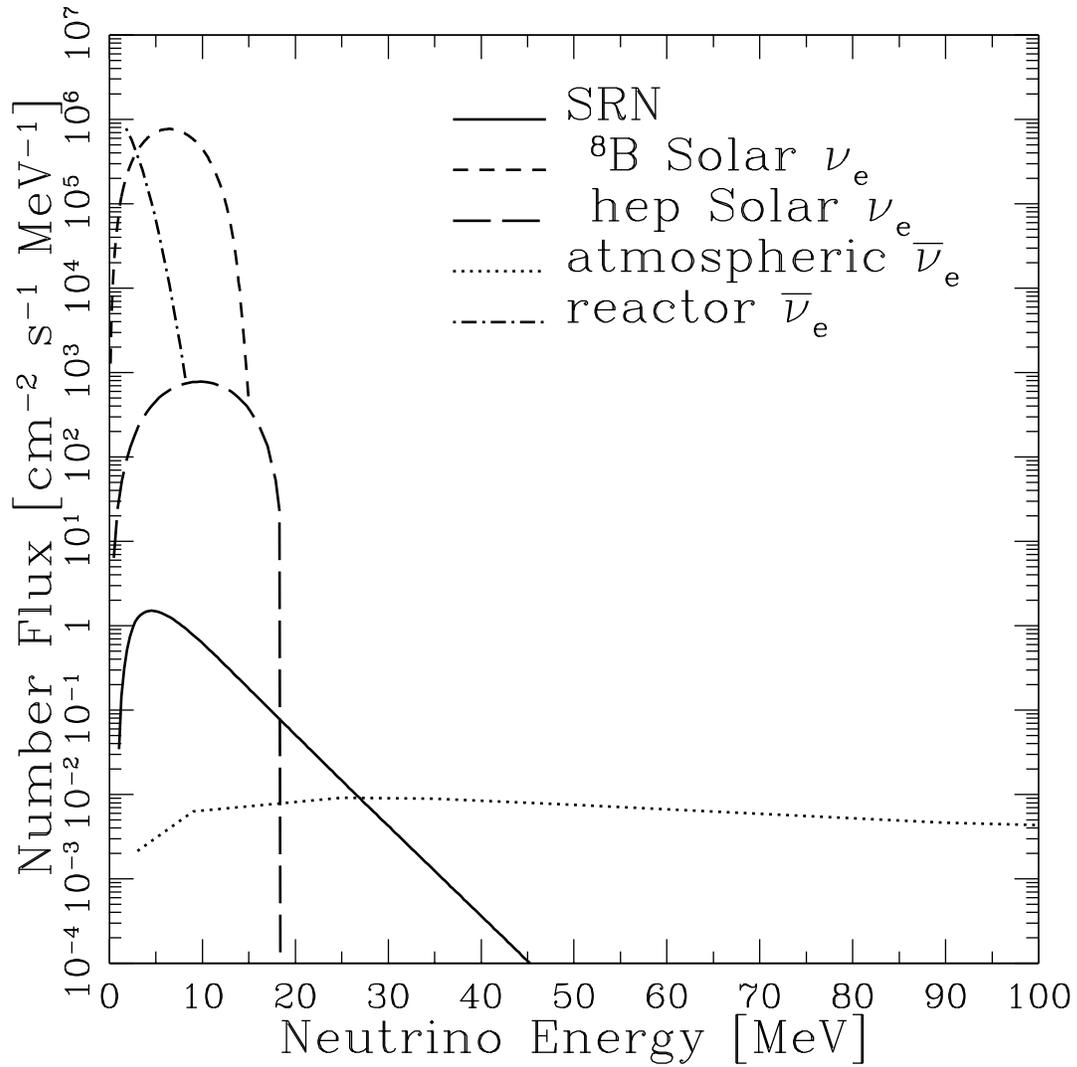}
 \caption{Number flux of SRN compared to other background neutrinos. 
 No oscillation and SN1 model are assumed for SRN flux.}  
\label{fig:background}
\end{center}
\end{figure}

We discuss in this subsection about neutrinos from other sources
which may become an obstacle to the SRN detection.  They are atmospheric
and solar neutrinos, anti-neutrinos from nuclear reactors, and decay
electrons from invisible muons.  We show in Fig. \ref{fig:background}
the number flux of SRN and these background events, and we discuss the
each background event below.

The flux of the atmospheric neutrinos is usually calculated using
Monte Carlo method including various relevant effects (flux of primary
cosmic rays, solar modulation, geomagnetic field, interaction of
cosmic rays in the air, and so on), and in that simulation
one-dimensional approximation is used, i.e., after the interaction of
primary cosmic ray particles with air nuclei, all the particles are 
assumed to be moving along the line of the momentum vector of the primary
cosmic ray particles \cite{rf:Kajita}. ( Recently
preliminary results of three-dimensional flux calculations have been
reported \cite{rf:Battistoni}.)  There are many authors who calculated
the atmospheric neutrino flux (see, e.g., \cite{rf:Lipari} for a
recent result).  We use in this paper the flux calculated by Gaisser
et al. \cite{rf:Gaisser} and Barr et al. \cite{rf:Barr}.  Although
their calculations are rather old, the fluxes of low energy neutrinos,
in which we are interested, are also calculated, while most of other
papers show only higher energy region ($>1~\mathrm{GeV}$).

Solar neutrino flux is dominant at energy range 10--20 MeV.  We use
the flux predicted by the standard solar model (SSM) in
Fig. \ref{fig:background} \cite{rf:Bahcall_2}.  Since the solar
neutrinos are not $\bar{\nu}_e$'s but $\nu_e$'s, the cross section for
solar neutrinos is about two order smaller than that for
$\bar{\nu}_e$'s.  Furthermore recoil electrons scattered by solar
neutrinos strongly concentrate to the opposite direction of the Sun,
in contrast to the isotropic distribution of $\bar{\nu}_e$ events.
Therefore the solar neutrino is an avoidable background.  In fact, we
can demonstrate how we can avoid the solar neutrino background, as
follows.  The solar neutrino event number is about $5\times 10^3$
$\mathrm{yr}^{-1}(22.5\mathrm{kton})^{-1}$ \cite{rf:Fukuda_1}, about 3
or 4 orders of magnitude larger than our expected event rate $\sim 1$
$\mathrm{yr}^{-1}(22.5\mathrm{kton})^{-1}$.  The recoil electrons
scatter obeying the Gaussian of one-sigma error $\sigma \sim 25^\circ$
for 10 MeV electrons \cite{rf:Nakahata}.  Then, 99.98\% of solar
neutrino events within $\sim 3.7\sigma$ (corresponding to $\sim
90^\circ$) can be avoided.  Therefore, restricting our discussion on
the nearby side of the hemisphere from the Sun, we can ignore the
solar neutrino events.

The third background which we must consider is anti-neutrinos from
nuclear reactors.  In each nuclear reactor, almost all the power comes
from the fissions of the four isotopes, ${}^{235}$U ($\sim 75\%$),
${}^{238}$U ($\sim 7\%$), ${}^{239}$Pu ($\sim 15\%$), and ${}^{241}$Pu
($\sim3\%$) \cite{rf:Bemporad}.  Each isotope produces a unique
electron anti-neutrino spectrum through the decay of its fission
fragments and their daughters. The $\bar{\nu}_e$ spectrum from
${}^{235}$U, ${}^{239}$Pu, and ${}^{241}$Pu can be derived using the
semi-empirical formula with which we fit data of detected
$\beta$-spectrum from fission by thermal neutrons \cite{rf:Hahn}.
(${}^{238}$U undergoes only fast neutron fission and hence electron
spectrum from ${}^{238}$U cannot by measured by this kind of
experiment.)  Above 7 MeV, the number of $\beta$ counts drops
dramatically and fitting error becomes large.  In addition, with this
method, as we determine the maximum $\beta$ energy and derives the
energy distribution below that energy, it is difficult to estimate the
errors on the high energy range \cite{rf:Suzuki_2}.  While the
$\bar{\nu}_e$ spectra in Ref. \cite{rf:Hahn} are given as tables, we
use for simplicity somewhat less accurate analytical approximation
given in Ref. \cite{rf:Vogel_2}.  As a normalization factor we use
energy-integrated $\bar{\nu}_e$ flux at Kamioka, $1.34 \times 10^6
~\mathrm{cm^{-2} s^{-1}}$, which are the summation of the flux from
various nuclear reactors in Japan and Korea \cite{rf:Bemporad}.  This
fit is not valid above $\sim 8$ MeV.  However, there is an
estimation that we would get few (10 or less) events per
year above 10 MeV from reactors\cite{rf:Suzuki_2}.

\begin{figure}[htbp]
\begin{center}
\includegraphics[width=15cm]{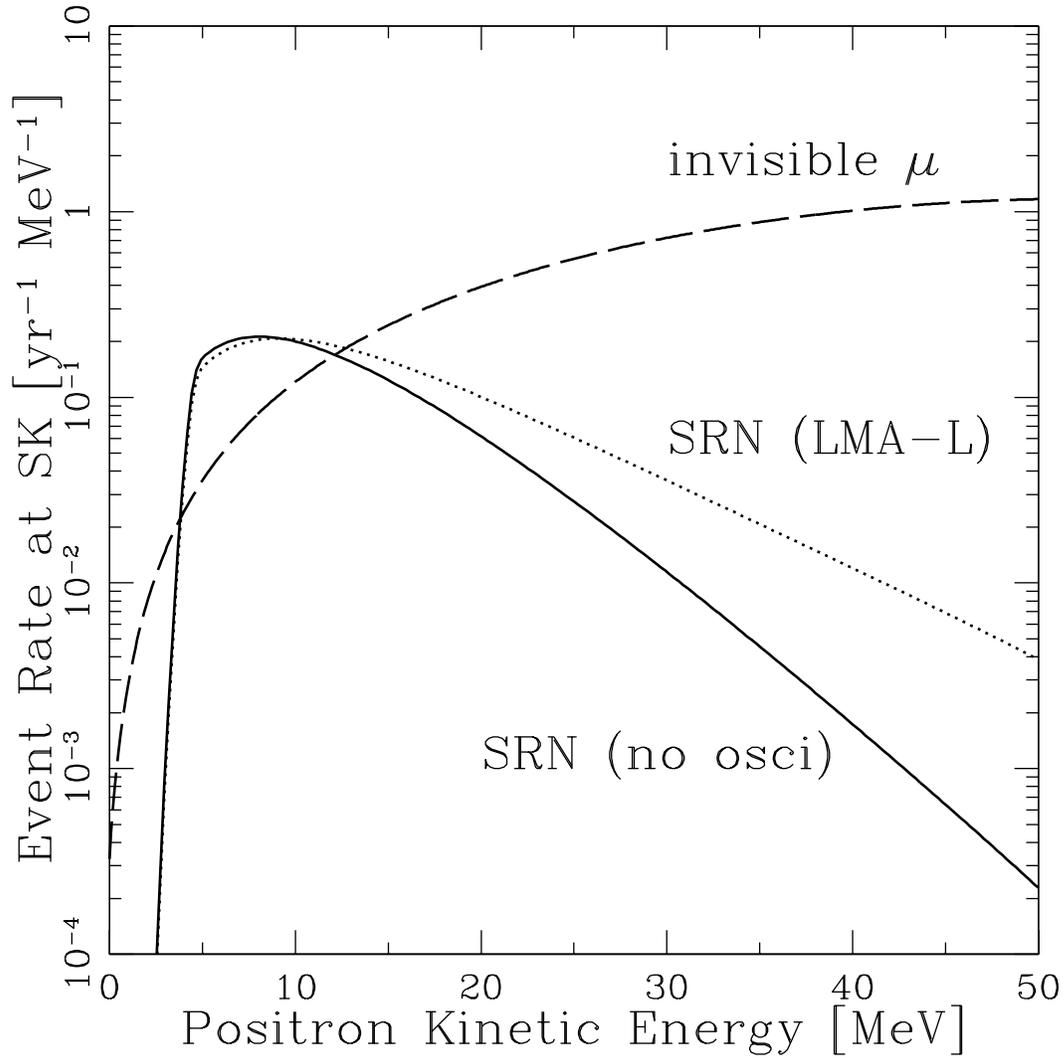}
 \caption{Event rate at SK detector of SRN and invisible $\mu$ decay
 products. Two oscillation models are shown (no oscillation and LMA-L)
 assuming SN1 model for supernova rate evolution.}  
\label{fig:invisible_mu}
\end{center}
\end{figure}

With three backgrounds we discussed above, we expect the energy window
of SRN events opening from 10 MeV to 27 MeV.  However, according to
Kaplighat et al. \cite{rf:Steigman} electrons or positrons from
invisible muons are the largest background in the energy window from
19 to 35 MeV.  This invisible muon event is illustrated as
follows. The atmospheric neutrinos produce muons by interaction with
the nucleons (both free and bound) in the fiducial volume.
If these muons are produced with energies below \v{C}herenkov
radiation threshold (kinetic energy less than 53 MeV), then they will
not be detected (``invisible muons''), but their decay-produced
electrons and positrons will be.  Since the muon decay signal will
mimic the $\bar{\nu}_e p \rightarrow n e^+$ process in SK, it is
difficult to distinguish SRN from these events.  The energy spectrum
of this invisible muon events is obtained by the stopped muon decay
spectrum
\begin{equation}
\frac{dN}{dE_e}=
\frac{G_F^2}{12\pi^3}m_\mu^2E_e^2\left(3-\frac{4E_e}{m_\mu}\right),
\end{equation}
(Michel spectrum \cite{rf:Michel}), where $G_F$ is the Fermi constant,
$m_\mu$ the muon mass, and $E_e$ the electron (positron) energy.  This
equation is valid for $E_e<m_\mu/2$.  From the observation of
Kamiokande II detector the estimated event rate from these muon decays
is around unity for 0.58 kton yr exposure and this is forming the
principal source of background after the various cuts had been
implemented \cite{rf:Zhang}.  Kaplinghat et al. \cite{rf:Steigman}
used this value corrected for SK by multiplying the volume ratio and
concluded that it is impossible to detect SRN unless their upper limit
realizes and maximum neutrino oscillation occurs.  This event rate is
now inferred directly from the new data of the SK, which is about 100
per 1,258 days per 22.5 kton fiducial volume above 18 MeV
\cite{rf:Suzuki_2,rf:Nakahata_2}.  This value corresponds to about
20\% of the Kamiokande II data.  In Fig. \ref{fig:invisible_mu} we
show SRN event rate compared to the invisible muon events.  From this
figure we conclude that SRN events can be seen only below about 12
MeV.

In practice, there is another serious background, i.e., spallation 
products induced by cosmic ray muons. Ultra high energy cosmic ray
muons spall oxygens in the detector, and radioactive decay processes
of these spalled nuclei occur. The event rate of the spallation 
background is several hundred per day per 22.5 kton. Although most of 
them can be rejected by the information of preceding muons, even a 
small fraction can not be. (Roughly, this spallation products produces
about 200,000 events per a year. Because expected SRN event rate is 
less than 1 per a year, we should reject all these 200,000 events.
For future detectors this problem is also quite difficult to solve.)
Then, this makes a serious background at the 
energy range below the maximum energy of beta spectrum of spallation 
products, 16 MeV \cite{rf:Nakahata_2,rf:Fukuda_3}. 
From these discussions presented 
in this subsection we conclude that there is no energy window of SRN.

\subsection{Calculating the detectable event rate at SK}

In the previous subsection we have found that there is no energy region
where SRN is dominant. However, we can detect the SRN events by subtracting
the other background events from total detected events. 
We consider the energy range $17<(T_e/ \mathrm{MeV})<25$, where
$T_e$ is positron kinetic energy.  This range corresponds to
$19<(E_{\bar{\nu}_e}/ \mathrm{MeV})<27$ by the simple relation,
$E_{\bar{\nu}_e}=T_e+1.8\mathrm{MeV}$. We find two advantages in using this 
energy region. First, SRN event rate is rather large, and second, the 
background (invisible muon) event rate is fairly well known by 
SK observation. We show in Table \ref{table:final} the SRN event rate at SK
in this energy range, i.e., $0.4-0.8 ~\mathrm{yr^{-1}}$.
In contrast, the event rate of the invisible muon over the same energy range
is $3.4~\mathrm{yr^{-1}}$.
When SRN event rate is larger than the statistical error of background event
rate, we can conclude that the SRN is detectable as a distortion of the 
expected invisible muon background event. Unfortunately, only one year 
observation does not provide any useful information about SRN. However, we can
expect that ten-year observation provides several statistically meaningful
results. The statistical error of invisible muon events in ten years is 
$\sqrt{34}=5.8$, which is smaller than the event rate of LMA models and is
larger than that of SMA and no oscillation models.
Then we conclude that these neutrino oscillation models can be distinguished
by the observation of the event rate of invisible muon events. (If there is
a discrepancy from expected event rate, this is due to SRN events and LMA 
models are favored.)

In future, it is expected that next generation of water {\v C}herenkov
detectors have much larger volume than that of SK.  For example
HyperKamiokande project is now under consideration.  HyperKamiokande detector
is planed to be a water {\v C}herenkov detector whose mass is about 1,000,000
tons (about 20 times larger than SK), and its location is near SK detector.
We expect that the SRN event becomes about 10 per one year for this detector,
and statistically sufficient discussion of SRN is possible even using 
only one year data.

\subsection{SNO and KamLAND detectors}

In this subsection we discuss SRN detectability with SNO \cite{rf:SNO_2} and KamLAND \cite{rf:KamLAND}.
An advantage of these detectors is that we are able to identify $\bar{\nu}_e$ events using delayed coincidence signals.
For this reason, we can remove other backgrounds from non-$\bar{\nu}_e$ origin (solar neutrinos, invisible muon decay products, and spallation products).

At SNO detector, used reaction for $\bar{\nu}_e$ detection is
\begin{equation}
\bar{\nu}_e + d \longrightarrow e^+ + n + n.
\label{eq:deuteron}
\end{equation}
Then these neutrons react with surrounding nuclei through
\begin{eqnarray}
n+d\longrightarrow {}^3\mathrm H + \gamma ~(E_\gamma=6.3\mathrm{MeV},
	{\rm efficiency}:24\%),\\
n+{}^{35}\mathrm{Cl} \longrightarrow {}^{36}\mathrm{Cl}+\gamma
	~(E_\gamma=8.6 \mathrm{MeV}, {\rm efficiency}:83\%),
\end{eqnarray}
where NaCl is added to efficiently capture neutrons.
This delayed signal with the preceding {\v C}herenkov radiation from $e^+$ in eq. (\ref{eq:deuteron}) shows that the detected neutrino is $\bar{\nu}_e$.
Using this criterion we can reject backgrounds from non-$\bar{\nu}_e$ origin.
These are solar neutrinos, invisible muon decay products, and spallation products, which are great obstacle for SRN detection at SK.
Thus, we can find the energy window below $T_e\simeq 23 \mathrm{MeV}$.
Actual calculation shows that the expected event rate at SNO is $\sim 0.03 ~\mathrm{yr}^{-1}$. (We used cross section calculated in Ref. \cite{rf:Haxton}.)

We can use a similar criterion at KamLAND.
$\bar{\nu}_e$'s are detected through below reactions:
\begin{equation}
\bar{\nu}_e + p \longrightarrow e^+ + n,
\end{equation}
\begin{equation}
n + p \longrightarrow d + \gamma ~(E_\gamma=2.2\mathrm{MeV}).
\end{equation}
The energy window for SRN is from 10 MeV to 25 MeV.
(Below 10 MeV reactor $\bar{\nu}_e$'s are large.)
In that range the calculated event rate is $\sim 0.1 ~\mathrm{yr}^{-1}$.

Unfortunately these values (0.03 yr$^{-1}$ for SNO; 0.1 yr$^{-1}$ for KamLAND) are quite small, since the fiducial volume of these detectors (1 kton) is much smaller than that of SK.
However, the future same kind of detector of larger volume, if ever built, might detect SRN.

\section{Acknowledgments}

We would like to thank A. Suzuki, M. Nakahata, and Y. Fukuda for useful discussions and also would like to thank K. Takahashi for preparing the numerical data of the neutrino oscillation and for useful discussions. 
S. Ando also would like to thank S. Nagataki for useful discussions.
This work was supported in part by
Grants-in-Aid for Scientific Research provided by the Ministry of Education,
Science and Culture of Japan through Research Grant No.07CE2002.



\clearpage

\clearpage

\begin{table}
\caption{Sets of mixing paremeter for calculation
	\label{table:parameter}}
\begin{center}
\begin{tabular}{ccccccc}\hline \hline
model  & $\sin^{2} 2 \theta_{12}$ & $\sin^{2} 2 \theta_{23}$ & $\sin^{2} 2 \theta_{13}$ 
& $\Delta m_{12}^{2}({\rm eV}^{2})$  & $\Delta m_{13}^{2}({\rm eV}^{2})$ 
& $\nu_{\odot}$ problem \\ \hline 
LMA-L &  0.87  & 1.0 & 0.043 & $7.0 \times 10^{-5}$ & $3.2 \times 10^{-3}$ & LMA \\
LMA-S &  0.87  & 1.0 & $1.0 \times 10^{-6}$ & $7.0 \times 10^{-5}$ & $3.2 \times 10^{-3}$ 
& LMA \\
SMA-L &  $5.0 \times 10^{-3}$  & 1.0 & 0.043  & $6.0 \times 10^{-6}$ & $3.2 \times 10^{-3}$ 
& SMA \\ 
SMA-S &  $5.0 \times 10^{-3}$  & 1.0 & $1.0 \times 10^{-6}$ & $6.0 \times 10^{-6}$ & $3.2 \times 10^{-3}$
& SMA \\ \hline
\end{tabular} 
\end{center}
\end{table}

\begin{table}
\caption{Number flux of SRN. Each entry is the number flux 
	in unit cm$^{-1}$ s$^{-1}$ for each supernova rate 
	and oscillation models.
	These fluxes are integrated over whole energy range.
	\label{table:flux}}
\begin{center}
\begin{tabular}{c|ccccc}\hline \hline
models & LMA-L & LMA-S & SMA-L & SMA-S & no oscillation \\ \hline
SN1 & 11.2 & 11.3 & 12.3 & 12.2 & 11.9 \\
SN2 & 12.2 & 12.2 & 13.4 & 13.2 & 12.9 \\
SN3 & 13.8 & 13.8 & 15.1 & 14.9 & 14.6 \\
\hline
\end{tabular} 
\end{center}
\end{table}

\begin{table}
\caption{SRN event rate at SK detector. Each entry is the event rate 
	in unit yr$^{-1}$ for each supernova rate and oscillation models.
	Integrated energy range is from 17 to 25 MeV.
	\label{table:final}}
\begin{center}
\begin{tabular}{c|ccccc}\hline \hline
models & LMA-L & LMA-S & SMA-L & SMA-S & no oscillation \\ \hline
SN1 & 0.73 & 0.72 & 0.46 & 0.45 & 0.44 \\
SN2 & 0.69 & 0.68 & 0.44 & 0.43 & 0.42 \\
SN3 & 0.76 & 0.75 & 0.50 & 0.49 & 0.48 \\
\hline 
\end{tabular} 
\end{center}
\end{table}

\end{document}